\title{Response of a DC RC circuit in series embedded with a rotating varco}
\author{
Sparisoma Viridi\thanks{dudung@fi.itb.ac.id}\\
Nuclear Physics and Biophysics Research Division\\
Institut Teknologi Bandung, Bandung 40132, Indonesia\\
\and
Siti Nurul Khotimah\thanks{nurul@fi.itb.ac.id}\\
Nuclear Physics and Biophysics Research Division\\
Institut Teknologi Bandung, Bandung 40132, Indonesia\\
}
\date{\today}

\documentclass[12pt]{article}
\usepackage{graphicx}
\usepackage{amssymb}
\usepackage{hyperref}

\begin{document}
\maketitle

\begin{abstract}
Response in electric potensial $U$ from a DC RC circuit in series has been obtained by varying the capacitance of the capacitor $C$. Variation of the capacitance begins at $t_0 > 0$. Alternating response is observed when the periode of change of capacitance in the same order as $\tau = RC$, which is held constant for $t < t_0$ and varied in triangle form for $t > t_0$.
\end{abstract}

\section{Introduction}

In a simple direct current circuit (DC) in series consisted of a battery, a resistor, and a capacitor current response is normally exponentially decayed to zero with increasing of time \cite{Halliday_2011}. Alternating current response can be obtained when the voltage source has a alternating current (AC) type \cite{Serway_2010}. An AC response in a DC resistor-capasitor (RC) circuit in series can be also observed when the capacitance of the capacitor is varied with time, as reported in this work.

\section{Variabel condensator}

A variable condensator or also known as varco is a capacitor that its capacitance can be varied. One type is varied by turning its rotor clockwise and counterclockwise \cite{Wikipedia_Varco}. Suppose that a varco consists of a pair of half circle plate with diameter $D$ then its capacitance can be defined as

\begin{equation}
\label{eq_varco_capacitance}
C = \frac{\epsilon_0 D^2}{8d} |\pi - \theta|,
\end{equation}

with $d$ is the distance between pair of the half circle plate and $\epsilon_0$ is vacuum permitivity. The angle $\theta$ is periodic in $[0, 2\pi]$. If the varco rotor is rotating with uniform angular velocity $\omega$ then the angular position is

\begin{equation}
\label{eq_varco_angular_position}
\theta = \omega t.
\end{equation}

Substitution of Equation (\ref{eq_varco_capacitance}) into Equation (\ref{eq_varco_angular_position}) will give the capacitance of the varco as function of time

\begin{equation}
\label{eq_varco_capacitance_time}
C = \frac{\epsilon_0 D^2}{8d} |\pi - \omega t|.
\end{equation}

\section{A DC RC circuit in series}

A DC RC circuit in series consisted of battery $\varepsilon$, resistor $R$, and capacitor $C$ has a relation

\begin{equation}
\label{eq_DC_RC_series}
\varepsilon - IR - \frac{Q}{C} = 0,
\end{equation}

with relation between current $I$ and charge $Q$ is

\begin{equation}
\label{eq_current_charge}
I = \frac{dQ}{dt}.
\end{equation}

Derivation of Equation (\ref{eq_DC_RC_series}) using (\ref{eq_current_charge}) will give the relation

\begin{equation}
\label{eq_DC_RC_series_current}
\frac{dI}{dt} + \frac{I}{\tau} - \frac{Q}{C\tau} \frac{dC}{dt} = 0,
\end{equation}

with

\begin{equation}
\label{eq_DC_RC_tau}
\tau = RC.
\end{equation}

Subtituting expression of $Q/C$ from Equation (\ref{eq_DC_RC_series}) into Equation (\ref{eq_DC_RC_series_current}) will give

\begin{equation}
\label{eq_DC_RC_series_current_2}
\frac{dI}{dt} = - \left(\frac{1}{\tau} + \frac{R}{\tau}\frac{dC}{dt}\right) I + \frac{\varepsilon}{\tau} \frac{dC}{dt},
\end{equation}

that will reduce to known result \cite{Halliday_2011} for constant capacitance, $dC/dt = 0$. In this case expression of $dC/dt$ can be found from Equation 

\begin{equation}
\label{eq_varco_capacitance_time_derivation}
\frac{dC}{dt} = \frac{\epsilon_0 D^2}{8d} \cdot \left\{
\begin{array}{lr}
0, & t < t_0 \\
-\omega, & t_0 + 2n\pi/\omega \le t < t_0 + (2n + 1)\pi/\omega \\
\omega, & t_0 + (2n + 1)\pi/\omega \le t < t_0 + 2(n + 1)\pi/\omega \\
\end{array}
\right.,
\end{equation}

with $t_0$ is start time of rotating varco. It is chosen that the capacitance

\begin{equation}
\label{eq_varco_capacitance_time_2}
C = \frac{\epsilon_0 D^2}{8d} \cdot \left\{
\begin{array}{lr}
\pi, & t < t_0 \\
\pi - \omega t, & t_0 + 2n\pi/\omega \le t < t_0 + (2n + 1)\pi/\omega \\
\omega t - \pi, & t_0 + (2n + 1)\pi/\omega \le t < t_0 + 2(n + 1)\pi/\omega \\
\end{array}
\right..
\end{equation}

Then it can also be obtained that

\begin{equation}
\label{eq_DC_RC_tau_2}
\tau = \frac{R\epsilon_0 D^2}{8d} \cdot \left\{
\begin{array}{lr}
\pi, & t < t_0 \\
\pi - \omega t, & t_0 + 2n\pi/\omega \le t < t_0 + (2n + 1)\pi/\omega \\
\omega t - \pi, & t_0 + (2n + 1)\pi/\omega \le t < t_0 + 2(n + 1)\pi/\omega \\
\end{array}
\right..
\end{equation}

\section{Numerical solution}

Equation (\ref{eq_DC_RC_series_current_2}) with Equation(\ref{eq_varco_capacitance_time_derivation})-(\ref{eq_DC_RC_tau_2}) can be solved numerically, using {\it e.g.} Euler method \cite{Wikipedia_Euler}, which is

\begin{equation}
\label{eq_Euler_I}
I(t + \Delta t) = I(t) +  \left(\frac{dI}{dt}\right) \Delta t.
\end{equation}

in this case. It is chosen that

\begin{equation}
\label{eq_Euler_Delta_t}
\Delta t = 10^{-3} \frac{2\pi}{\omega}.
\end{equation}

\section{Results and discussion}

Following parameters are used in the calculation: $R= 1 ~{\rm k}\Omega$, $\varepsilon = 10~{\rm V}$, $t_0 = 2~\mu{\rm s}$, $D = 0.5~{\rm m}$, $d = 1~{\rm mm}$, and $\Delta t$ = 10 ns. Frequency, $f = \omega/2\pi$, is varied: 10 Hz - 1 MHz. It is found that $C(t < t_0)$ = 0.869258 nF and $\tau(t < t_0)$ = 0.869258 ms.

\begin{figure}[h]
\centering
\includegraphics[width=13.5cm]{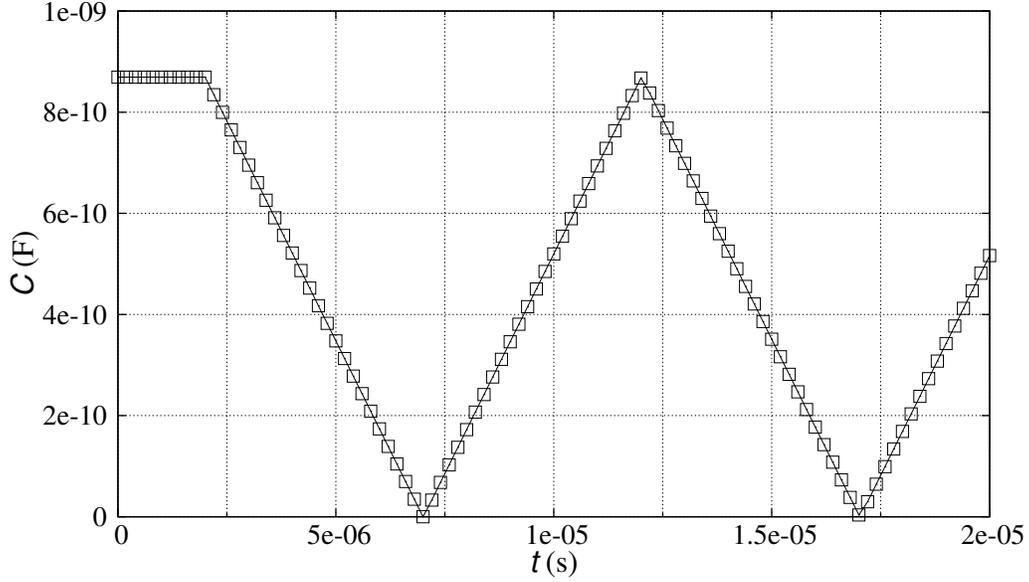}
\caption{\label{fg_C_f1E5}. A typical form of $C$ for $f = 10^5$ Hz.}
\end{figure}

\begin{figure}[h]
\centering
\includegraphics[width=13.5cm]{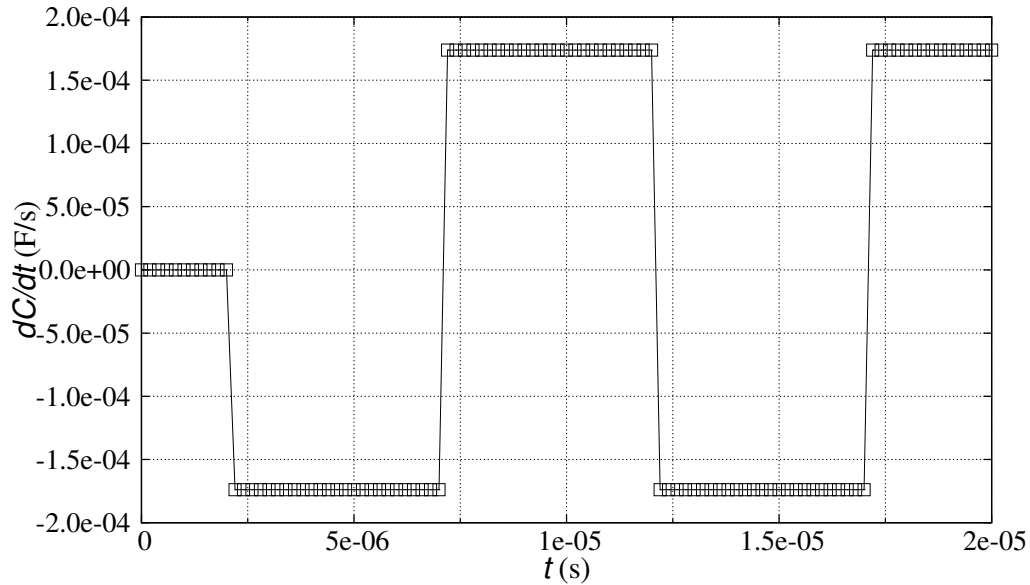}
\caption{\label{fg_dCdt_f1E5}. A typical form of $dC/dt$ for $f = 10^5$ Hz.}
\end{figure}

\begin{figure}[h]
\centering
\includegraphics[width=13.5cm]{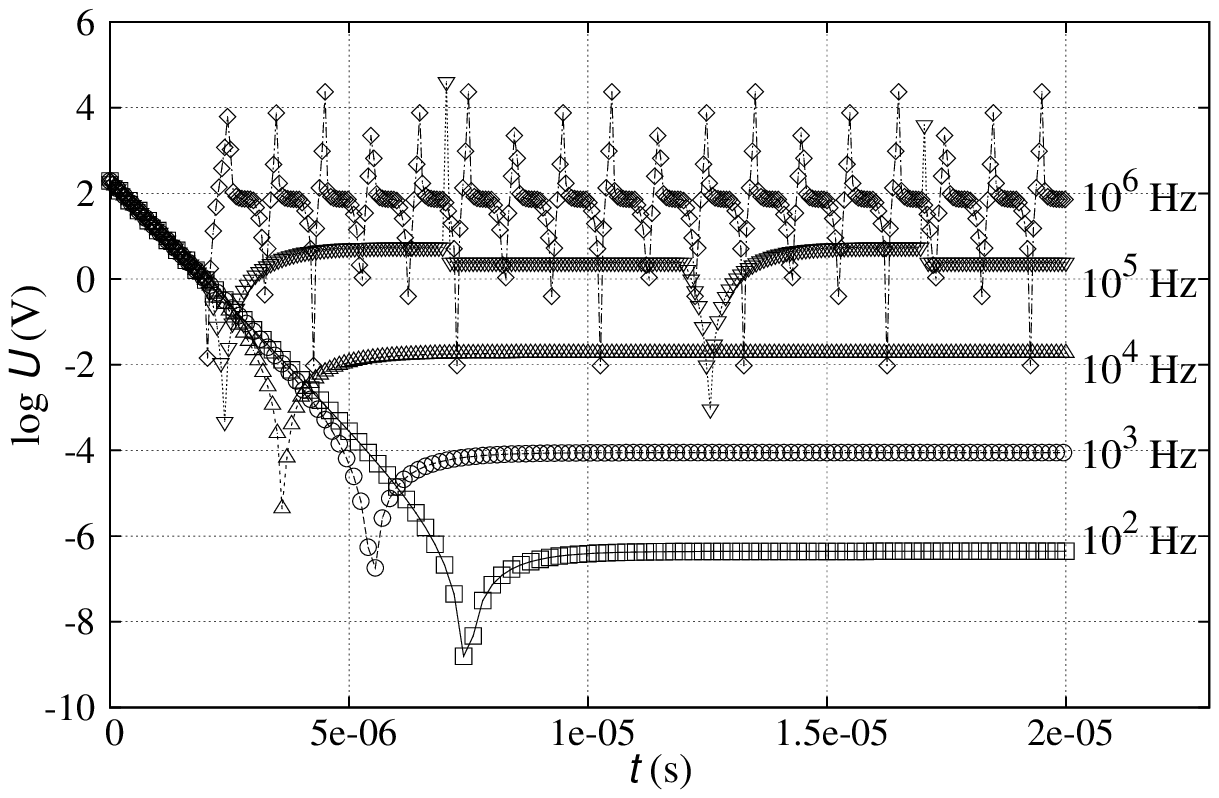}
\caption{\label{fg_logU_t}. Response of log $U$ against time $t$ for several frequency $f$ = 10$^2$, 10$^3$, 10$^4$, 10$^5$, and 10$^6$ Hz.}
\end{figure}

Response in form of $U = IR$ is given in Figure \ref{fg_logU_t}, which illustrates that only periode of rotating varco that nearly the same order as $\tau$ could give the alternating response as $f = 10^5$ and $10^6$ Hz.

In the calculation detail an exception must be implemented to avoid value of $\tau = 0$ since it is allowed as given by $C = 0$ as illustrated in Figure \ref{fg_C_f1E5} and defined in Equation (\ref{eq_DC_RC_tau_2}).

\section{Conclusion}

An alternating response can be obtained from a DC RC circuit in series by varying the value of capacitance of the capacitor which is assume to be a varco. The variation is taken place by rotating the rotor of the varco with a constant angular velocity. Only rotation periode that has the same order as $\tau = RC$ will give an alternating response.

\bigskip\bigskip\noindent
{\bf \Large Acknowledgements}\\ \\
Authors would like to thank to Institut Teknologi Bandung Alumni Association Research Grant in year 2011 for partially supporting to this work.

\bibliographystyle{unsrt}
\bibliography{manuscript}

\begin{thebibliography}{1}

\bibitem{Halliday_2011}
David Halliday, Robert Resnick, and Jearl Walker.
\newblock {\em Fundamentals of Physics}.
\newblock John Wiley \& Sons, 9th edition, 2011.

\bibitem{Serway_2010}
Raymond~A. Serway and Jr. John W.~Jewett.
\newblock {\em Physics for Scientists and Engineers with Modern Physics}.
\newblock Brooks/Cole, 8th edition, 2010.

\bibitem{Wikipedia_Varco}
Wikipedia contributors.
\newblock Variabel capacitor.
\newblock {\em Wikipedia, The Free Encyclopedia}, oldid:470996212, [accessed 31
  Jan 2011].

\bibitem{Wikipedia_Euler}
Wikipedia contributors.
\newblock Euler method.
\newblock {\em Wikipedia, The Free Encyclopedia}, oldid:467854539, [accessed 31
  Jan 2011].

\end{thebibliography}

\end{document}